\documentclass[aps,prl,preprint,nopacs,superscriptaddress]{revtex4}
\usepackage{amsmath}
\usepackage{amssymb}
\usepackage{graphicx}
\usepackage{hyperref}
\usepackage[utf8]{inputenc}
\usepackage{bbm} 
\newcommand{\beq}{\begin{equation}}
\newcommand{\eeq}{\end{equation}}
\newcommand{\beqs}{\begin{equation*}}
\newcommand{\eeqs}{\end{equation*}}
\newcommand{\s}{Co$_2$MnGa}
\newcommand{\eb}{E_\textrm{B}}

\newcommand{\ef}{E_\textrm{F}}
\newcommand{\ab}{{\it ab initio}}
\newcommand{\Ab}{{\it Ab initio}}
\newcommand{\exda}{Extended Data Fig. }
\newcommand{\exdas}{Extended Data Figs. }

\newcommand{\pana}{a}
\newcommand{\panb}{b}
\newcommand{\panc}{c}
\newcommand{\pand}{d}
\newcommand{\pane}{e}
\newcommand{\panf}{f}

\newcommand{\cpana}{{\bf a}}
\newcommand{\cpanb}{{\bf b}}
\newcommand{\cpanc}{{\bf c}}
\newcommand{\cpand}{{\bf d}}
\newcommand{\cpane}{{\bf e}}
\newcommand{\cpanf}{{\bf f}}

\begin{document}

\title{Observation of a linked loop quantum state}

\author{Ilya Belopolski\footnote{These authors contributed equally to this work.}}\email{ilya.belopolski@riken.jp}
\affiliation{Laboratory for Topological Quantum Matter and Spectroscopy (B7), Department of Physics, Princeton University, Princeton, New Jersey 08544, USA}
\affiliation{RIKEN Center for Emergent Matter Science (CEMS), Wako, Saitama 351-0198, Japan}

\author{Guoqing Chang$^*$}
\affiliation{Division of Physics and Applied Physics, School of Physical and Mathematical Sciences, Nanyang Technological University, 21 Nanyang Link, 637371, Singapore}

\author{Tyler A. Cochran$^*$}
\affiliation{Laboratory for Topological Quantum Matter and Spectroscopy (B7), Department of Physics, Princeton University, Princeton, New Jersey 08544, USA}

\author{Zi-Jia Cheng$^*$}
\affiliation{Laboratory for Topological Quantum Matter and Spectroscopy (B7), Department of Physics, Princeton University, Princeton, New Jersey 08544, USA}

\author{Xian P. Yang}
\affiliation{Laboratory for Topological Quantum Matter and Spectroscopy (B7), Department of Physics, Princeton University, Princeton, New Jersey 08544, USA}

\author{Cole Hugelmeyer}
\affiliation{Department of Mathematics, Princeton University, Princeton, New Jersey 08544, USA}

\author{Kaustuv Manna}
\affiliation{Max Planck Institute for Chemical Physics of Solids, N\"othnitzer Stra{\ss}e 40, 01187 Dresden, Germany}
\affiliation{Department of Physics, Indian Institute of Technology Delhi, Hauz Khas, New Delhi 110016, India}

\author{Jia-Xin Yin}
\affiliation{Laboratory for Topological Quantum Matter and Spectroscopy (B7), Department of Physics, Princeton University, Princeton, New Jersey 08544, USA}

\author{Guangming Cheng}
\affiliation{Princeton Institute for Science and Technology of Materials, Princeton University, Princeton, New Jersey, 08544, USA}

\author{Daniel Multer}
\affiliation{Laboratory for Topological Quantum Matter and Spectroscopy (B7), Department of Physics, Princeton University, Princeton, New Jersey 08544, USA}

\author{Maksim Litskevich}
\affiliation{Laboratory for Topological Quantum Matter and Spectroscopy (B7), Department of Physics, Princeton University, Princeton, New Jersey 08544, USA}

\author{Nana Shumiya}
\affiliation{Laboratory for Topological Quantum Matter and Spectroscopy (B7), Department of Physics, Princeton University, Princeton, New Jersey 08544, USA}

\author{Songtian S. Zhang}
\affiliation{Laboratory for Topological Quantum Matter and Spectroscopy (B7), Department of Physics, Princeton University, Princeton, New Jersey 08544, USA}

\author{Chandra Shekhar}
\affiliation{Max Planck Institute for Chemical Physics of Solids, N\"othnitzer Stra{\ss}e 40, 01187 Dresden, Germany}

\author{Niels B. M. Schr\"oter}
\affiliation{Swiss Light Source, Paul Scherrer Institut, CH-5232 Villigen, Switzerland}

\author{Alla Chikina}
\affiliation{Swiss Light Source, Paul Scherrer Institut, CH-5232 Villigen, Switzerland}

\author{Craig Polley}
\affiliation{MAX IV Laboratory, Lund University, P.O. Box 118, 221 00 Lund, Sweden}

\author{Balasubramanian Thiagarajan}
\affiliation{MAX IV Laboratory, Lund University, P.O. Box 118, 221 00 Lund, Sweden}

\author{Mats Leandersson}
\affiliation{MAX IV Laboratory, Lund University, P.O. Box 118, 221 00 Lund, Sweden}

\author{Johan Adell}
\affiliation{MAX IV Laboratory, Lund University, P.O. Box 118, 221 00 Lund, Sweden}

\author{Shin-Ming Huang}
\affiliation{Department of Physics, National Sun Yat-Sen University, Kaohsiung 804, Taiwan}

\author{Nan Yao}
\affiliation{Princeton Institute for Science and Technology of Materials, Princeton University, Princeton, New Jersey, 08544, USA}

\author{Vladimir N. Strocov}
\affiliation{Swiss Light Source, Paul Scherrer Institut, CH-5232 Villigen, Switzerland}

\author{Claudia Felser}
\affiliation{Max Planck Institute for Chemical Physics of Solids, N\"othnitzer Stra{\ss}e 40, 01187 Dresden, Germany}

\author{M. Zahid Hasan} \email{mzhasan@princeton.edu}
\affiliation{Laboratory for Topological Quantum Matter and Spectroscopy (B7), Department of Physics, Princeton University, Princeton, New Jersey 08544, USA}
\affiliation{Princeton Institute for Science and Technology of Materials, Princeton University, Princeton, New Jersey, 08544, USA}
\affiliation{Materials Sciences Division, Lawrence Berkeley National Laboratory, Berkeley, CA 94720, USA}


\begin{abstract}
Quantum phases can be classified by topological invariants, which take on discrete values capturing global information about the quantum state \cite{news_Buchanan,GeoTopoPhys_Nakahara,Topo_ChaikinLubensky,Nobel_Haldane,RevMod_Topology_Wen,HeliumDroplet_Volovik, HeliumRMP_Volovik,TopoMag_Hoffmann,SkyrmionsNatNano_Tokura,news_ChernQHE_Avron,Colloquium_Zahid,Topo_Andrei,ReviewSSH_Kane,RMPTopoBandThy_Bansil,ReviewMagTopo_Tokura,ARCMP_me,Review_ZahidSuyangGuang,RMPWeylDirac_Armitage,RMPClassification_ChingKai}. Over the past decades, these invariants have come to play a central role in describing matter, providing the foundation for understanding superfluids \cite{HeliumDroplet_Volovik, HeliumRMP_Volovik}, magnets \cite{TopoMag_Hoffmann,SkyrmionsNatNano_Tokura}, the quantum Hall effect \cite{Nobel_Haldane,news_ChernQHE_Avron}, topological insulators \cite{Colloquium_Zahid,Topo_Andrei,ReviewSSH_Kane,RMPTopoBandThy_Bansil,ReviewMagTopo_Tokura}, Weyl semimetals \cite{ARCMP_me,Review_ZahidSuyangGuang,RMPWeylDirac_Armitage,RMPClassification_ChingKai} and other phenomena. Here we report a remarkable linking number (knot theory) invariant associated with loops of electronic band crossings in a mirror-symmetric ferromagnet \cite{Co2MnGa_me,Co2MnGa_Guoqing,NonAbelianLinks_WuSoluyanov,NodalLink_ZhongboYan,Knots_Ezawa,WeylLink_Yee,PentagonC_ShengbaiZhang}. Using state-of-the-art spectroscopic methods, we directly observe three intertwined degeneracy loops in the material's bulk Brillouin zone three-torus, $\mathbb{T}^3$. We find that each loop links each other loop twice. Through systematic spectroscopic investigation of this linked loop quantum state, we explicitly draw its link diagram and conclude, in analogy with knot theory, that it exhibits linking number $(2,2,2)$, providing a direct determination of the invariant structure from the experimental data. On the surface of our samples, we further predict and observe Seifert boundary states protected by the bulk linked loops, suggestive of a remarkable Seifert bulk-boundary correspondence. Our observation of a quantum loop link motivates the application of knot theory to the exploration of exotic properties of quantum matter.
\end{abstract}

\date{\today}
\maketitle

Quantum topology is powerful in understanding condensed matter systems that exhibit a winding \cite{news_Buchanan,GeoTopoPhys_Nakahara,Topo_ChaikinLubensky,Nobel_Haldane,RevMod_Topology_Wen,HeliumDroplet_Volovik, HeliumRMP_Volovik,TopoMag_Hoffmann,SkyrmionsNatNano_Tokura,news_ChernQHE_Avron,Colloquium_Zahid,Topo_Andrei,ReviewSSH_Kane,RMPTopoBandThy_Bansil,ReviewMagTopo_Tokura,ARCMP_me,Review_ZahidSuyangGuang,RMPWeylDirac_Armitage,RMPClassification_ChingKai}. Often, this winding occurs in real space. For example, in a magnetic material, the local magnetization may exhibit a rotating pattern centered around a point in real space, forming a magnetic vortex encoding an integer winding number \cite{Topo_ChaikinLubensky,TopoMag_Hoffmann}. Alternatively, the winding may occur in momentum space. For example, in a one-dimensional topological insulator, the quantum-mechanical wavefunctions wind as the momentum scans through the Brillouin zone \cite{Nobel_Haldane,RevMod_Topology_Wen,news_ChernQHE_Avron,Colloquium_Zahid,Topo_Andrei,ReviewSSH_Kane,RMPTopoBandThy_Bansil}. These two broad paradigms---order parameters, such as magnetization, which wind in real space and quantum wavefunctions which wind in momentum space---capture a vast landscape of topological phases of matter, spanning decades of research by myriad communities of physicists. Real-space order parameter winding further encompasses disclinations in liquid crystals; vortices in superconductors and superfluid $^4$He; and magnetic skyrmions, whose invariants are proposed as the basis for next-generation computing memory and logic \cite{GeoTopoPhys_Nakahara,Topo_ChaikinLubensky,HeliumDroplet_Volovik, HeliumRMP_Volovik,TopoMag_Hoffmann,SkyrmionsNatNano_Tokura}. On the other hand, momentum-space wavefunction winding is associated with emergent Dirac fermions in two- and three-dimensional topological insulators; Weyl fermions in topological semimetals; and the quantum Hall effect, which sets the prevailing von Klitzing standard of electrical resistance \cite{news_ChernQHE_Avron,Colloquium_Zahid,Topo_Andrei,ReviewSSH_Kane,RMPTopoBandThy_Bansil,ReviewMagTopo_Tokura,ARCMP_me,Review_ZahidSuyangGuang,RMPWeylDirac_Armitage,RMPClassification_ChingKai}. Despite their importance in modern physics, there is no indication that these two paradigms are exhaustive. Novel paradigms for topology promise to deepen our fundamental understanding of nature, as well as enable new quantum technologies.

Recently, there has been considerable interest in node loops, an electronic structure where two bands cross along a closed curve in momentum space \cite{Co2MnGa_me,Co2MnGa_Guoqing,PbTaSe2_Guang,MagSpaceGroups_Vishwanath,WeylLoopSuperconductor_Nandkishore,WeylLines_Kane,WeylDiracLoop_Nandkishore,RMPClassification_ChingKai}. Away from the crossing curve, the two bands disperse linearly, so that the node loop consists of a cone dispersion persisting along a loop. Within the paradigm of momentum-space wavefunction winding, node loops are topological, with a quantized Berry phase invariant \cite{Colloquium_Zahid,Topo_Andrei,ReviewSSH_Kane,RMPTopoBandThy_Bansil,WeylLines_Kane,WeylDiracLoop_Nandkishore,RMPClassification_ChingKai}. However, in contrast to other electronic structures studied to date \cite{news_ChernQHE_Avron,Colloquium_Zahid,Topo_Andrei,ReviewSSH_Kane,RMPTopoBandThy_Bansil,ReviewMagTopo_Tokura,ARCMP_me,Review_ZahidSuyangGuang,RMPWeylDirac_Armitage,RMPClassification_ChingKai}, node loops can link each other, encoding a linking number invariant (Fig. \ref{Fig1}\pana, \exda \ref{Fig0}) \cite{NonAbelianLinks_WuSoluyanov,NodalLink_ZhongboYan,Knots_Ezawa,WeylLink_Yee,PentagonC_ShengbaiZhang}. Unlike the traditional paradigms of winding, this linking number is associated with the composite loop structure of quantum-mechanical band crossings of the Hamiltonian. Such linked node loops offer the possibility of a new bridge between physics and knot theory. It has further been proposed that these links are governed by emergent non-Abelian node loop charges \cite{NonAbelianLinks_WuSoluyanov} and that the linking number determines the $\theta$ angle of the axion Lagrangian in certain node loop phases \cite{WeylLink_Yee, ShouChengZhang_HelixNodalLine, Biao_WilsonLoops}. Since the three-dimensional condensed matter Brillouin zone is a three-torus $\mathbb{T}^3$, linked node loops also offer the rare possibility of observing links in a space other than ordinary infinite space $\mathbb{R}^3$. Moreover, the Seifert surface of the bulk link is associated with topological boundary states, opening the possibility of a unique Seifert bulk-boundary correspondence in quantum matter \cite{Seifert_1935, NonHermitian_RMP, Seifert_LinhuLi, NonHermitian_Carlstrom, NonHermitian_Tidal}.

Ferromagnets with crystalline mirror symmetry naturally give rise to node loops. In this scenario, the ferromagnetic exchange interaction produces spin-split electronic bands which are generically singly-degenerate throughout momentum space, while mirror symmetry protects two-fold band degeneracies along closed curves confined to the momentum-space mirror planes \cite{MagSpaceGroups_Vishwanath}. Such node loops are called Weyl loops, by analogy with the two-fold degeneracy of a Weyl point \cite{ARCMP_me,Review_ZahidSuyangGuang,RMPWeylDirac_Armitage,WeylLoopSuperconductor_Nandkishore,WeylLines_Kane,WeylDiracLoop_Nandkishore}. Weyl loops are extremely effective at concentrating Berry curvature, giving rise to giant anomalous Hall and Nernst effects, up to room temperature and promising for technological applications \cite{Co2MnGa_me,Fe3Ga_Nakatsuji,Co2MnGa_Nakatsuji,Co2MnGa_Felser_Nernst,Co2MnGa_Thomas_Nernst,CrPt3_Anastasios}. In crystallographic space groups with multiple perpendicular mirror planes, different Weyl loops living in different mirror planes can naturally link each other \cite{Co2MnGa_Guoqing,PentagonC_ShengbaiZhang}. The ferromagnet \s\ exhibits a crystal structure with multiple perpendicular mirror planes and was recently observed to host electronic Weyl loops \cite{Co2MnGa_me,Co2MnGa_Guoqing}, bringing together the key ingredients for node loop links.

\s\ crystallizes in the full Heusler structure, with face-centered cubic Bravais lattice; space group $Fm\bar{3}m$ (No. 225); octahedral point group $O_h$; and conventional unit cell lattice constant $c = 5.77$ \AA\ (Fig. \ref{Fig1}\panb, \exda \ref{EFigCrys}). We observe that our \s\ samples are ferromagnetic, with Curie temperature $T_{\rm C} = 690$ K, consistent with earlier reports \cite{Co2MnGa_neutron,Co2MnGa_CurieTemp}. The point group includes mirror planes normal to $\hat{x}$, $\hat{y}$ and $\hat{z}$ (defined by the conventional unit cell lattice vectors, representative mirror plane shown in orange in \exda \ref{EFigCrys}\pana), as well as three $C_4$ rotation symmetries relating any one of these mirror planes to the other two. We first perform a characterization by atomic-level energy-dispersive X-ray spectroscopy (EDS), providing direct structural evidence that our \s\ samples are crystallographically well-ordered, show the expected lattice constant and exhibit these mirror and rotation symmetries (Fig. \ref{Fig1}\panb). The real-space mirror planes give rise to momentum-space mirror planes, labelled $M_1$ (normal to $\hat{z}$); $M_2$ ($\hat{y}$); and $M_3$ ($\hat{x}$, Fig. \ref{Fig1}\panc). These mirror planes contain the time-reversal invariant momenta $X_1$, $X_2$ and $X_3$, sitting at the centers of the square faces of the bulk Brillouin zone.

Motivated by the observation of mirror-symmetry-protected magnetic Weyl loops in \s\ \cite{Co2MnGa_me, Co2MnGa_Guoqing}, we explore the electronic structure of our samples on $M_1$, $M_2$ and $M_3$. We perform \ab\ calculations of \s\ in the ferromagnetic state, focusing on these three mutually-perpendicular mirror planes. We find that each mirror plane hosts a Weyl loop, and that the three Weyl loops link one another (Fig. \ref{Fig1}c, \exda \ref{EFig4}). To experimentally investigate this \ab\ prediction, we carry out angle-resolved photoemission spectroscopy (ARPES) using soft X-ray photons, optimized for exploring bulk electronic states \cite{ADRESS1_Strocov, ADRESS2_Strocov}. Without loss of generality, we consider the crystal cleaving plane in our photoemission experiments to be parallel to $M_1$. We first acquire a Fermi surface at a fixed incident photon energy $h\nu = 544$ eV, chosen to fix the $k_z$ momentum to this `in-plane' mirror plane ($M_1$, Fig. \ref{Fig1}\pand). We observe a diamond-shaped contour centered on $X_1$, which traces out a momentum-space trajectory encircling the square top face of the bulk Brillouin zone. We also observe a small circular feature at the corners of the Fermi surface, which arises from an unrelated band at $\Gamma$, irrelevant for what follows. We next perform a photon-energy dependence on the same sample, measuring from $h\nu = 500$ to $800$ eV, which allows us to access the electronic structure on the `out-of-plane' mirror $M_2$ (Fig. \ref{Fig1}\pane). We again observe a diamond-shaped loop contour, now centered on $X_2$ and encircling the square side face of the bulk Brillouin zone. We then rotate the sample by $90^{\circ}$ and repeat the same photon-energy dependence to capture the electronic structure on the other `out-of-plane' mirror $M_3$. We observe again a similar diamond-shaped loop contour centered on $X_3$ (Fig. \ref{Fig1}\panf). These systematic observations suggest a family of symmetry-related diamond-shaped loop contours, consistent with the octahedral point group $O_h$. Each of the three loops is related to the others by rotation symmetry and each lives in one of the three mirror planes.

To further understand the loop electronic structures, we examine energy-momentum photoemission spectra slicing through the $M_1$ loop (Fig. \ref{Fig2}\pana). We observe two bands which disperse toward each other and meet near the Fermi level, $\ef$ ($\eb = 0$), suggesting a cone dispersion. The presence of a cone dispersion in both slices further shows that the cone persists as we move in momentum space, following the $M_1$ loop Fermi surface. Since \s\ is ferromagnetic, we expect generically singly-degenerate bands throughout the Brillouin zone \cite{Co2MnGa_neutron,Co2MnGa_CurieTemp}. This suggests a cone dispersion consisting of singly-degenerate branches which exhibit a double degeneracy at the touching points, indicating a Weyl loop electronic structure. To better understand the cone dispersions, we perform \ab\ calculations of the electronic structure on energy-momentum slices corresponding to these ARPES spectra. The calculation shows a Weyl cone with characteristic two-fold degenerate crossing and linear dispersion away from the crossing (Fig. \ref{Fig2}\panb). Assembled together, this series of Weyl cones forms a Weyl loop. The ARPES and \ab\ calculations are in good agreement, further suggesting that we have observed a magnetic Weyl loop on $M_1$.

To characterize this Weyl loop using ARPES, we systematically track all cone crossings observed along the full $M_1$ loop trajectory (Fig. \ref{Fig2}\panc). We then fit the crossings to a two-band effective $k \cdot p$ Hamiltonian for a Weyl loop,
\beq
H = \sum_{k,\ a,b \in \{\pm\}} c^{\dag}_{ka} h_{ab}(k) c_{kb}, \hspace{1cm} h (k) = f(k)\sigma_z + v_{\rm F} q_z\sigma_x + g(k)\mathbbm{1}
\eeq
Here the $c^{\dag}_{ka}$ are fermionic creation operators, $k$ is the crystal momentum, $\sigma_z$ and $\sigma_x$ are Pauli matrices, $\mathbbm{1}$ is the $2 \times 2$ identity and $q_z \equiv k_z - 2\pi/c$ is the $\hat{z}$ component of the momentum measured relative to $M_1$, where $c$ is the conventional unit cell lattice constant. This Hamiltonian exhibits a Weyl loop on $q_z = 0$ with trajectory given by $f(k) = 0$, formed from two bands with opposite mirror eigenvalues. From our ARPES spectra, we experimentally extract the full Weyl loop trajectory by fitting to a low-order expansion around $X_1$, consistent with the symmetries of the system,
\beq
\label{traj}
f(k) = \gamma \big( 1 + \alpha (k_x^2 + k_y^2) + \beta k_x^2 k_y^2 \big)
\eeq
Here $\alpha$ and $\beta$ fix the Weyl loop trajectory and the scaling factor $\gamma$ sets an energy scale. The train of crossing points observed in ARPES is well-captured by $\alpha = -1.23 \pm 0.03$ \AA$^2$ and $\beta = -31.5 \pm 4.1$ \AA$^4$ (Fig. \ref{Fig2}\panc). We also find that our ARPES-extracted Weyl loop trajectory agrees well with the trajectory observed in \ab\ calculations (\exda \ref{EFig1}). The energy dispersion of the Weyl loop is set by $g(k)$, well-described by $g(k) = \delta + \eta \cos (4 \theta)$, where $\delta = -75 \pm 17$ meV, $\eta = 46 \pm 17$ meV, and $\theta$ is the ordinary polar angle of $k$, $\tan \theta \equiv k_y/k_x$. Away from the loop, the bands disperse linearly. To capture this dispersion, we examine a deeper constant-energy photoemission slice intersecting the Weyl loop (Fig. \ref{Fig2}\pand). We observe that the experimental dispersion is well-captured by $\gamma = 0.15 \pm 0.05$ eV. The extracted dispersion reaches the Fermi level within experimental resolution, suggesting that the observed Weyl loops are relevant for transport and other low-energy response. This result is consistent with previous reports that Weyl loops play a dominant role in the giant anomalous Hall effect and other exotic transport properties of \s\ \cite{Co2MnGa_me,Co2MnGa_Nakatsuji,Co2MnGa_Felser_Nernst,Co2MnGa_Thomas_Nernst}. Our analytical model, with all parameters extracted from photoemission spectra, allows us to achieve a complete experimental characterization of both the momentum trajectory and energy dispersion of the family of Weyl loops (Fig. \ref{Fig2}\pane, \exda \ref{EFig1}). 

We next investigate the composite structure formed by the three Weyl loops, focusing on the Weyl loop crossing itself (degeneracy curve, shown in cyan in Fig. \ref{Fig2}\pane). The loop crossing disperses in energy, so a constant-energy slice through the electronic structure typically intersects a loop crossing only at several discrete points (for example, the cyan dots in Fig. \ref{Fig2}\pand). However, despite this energy dispersion, we find that the typical `radius' of the Weyl loop, $|k|_{\rm avg} = 0.66$ \AA$^{-1}$, is much larger than the typical momentum separation of the two branches of the Weyl cone at the Fermi level, $\eta/v_{\rm F} = 0.07$ \AA$^{-1}$. Since $|k|_{\rm avg} \gg \eta/v_{\rm F}$, we can treat the Weyl loop crossing as approximately flat in energy and we can accurately capture its trajectory by examining a constant-energy slice near the Fermi level. Therefore, to understand the composite structure of the Weyl loops, we can focus on the $M_1$, $M_2$ and $M_3$ Weyl loop Fermi surfaces. We first consider the $M_1$ and $M_2$ Weyl loops in an extended zone scheme. We study two adjacent Brillouin zones (Fig. \ref{Fig3}\pana, inset) and zoom in on the momentum-space region around $X_1$. By plotting the $M_1$ and $M_2$ Weyl loop Fermi surfaces simultaneously in this region of three-dimensional momentum space, we observe that these two loops appear to link each other twice (Fig. \ref{Fig3}\pana). We next consider the $M_2$ and $M_3$ Weyl loops and we shift our momentum-space field of view to a region around $X_2$. We plot the $M_2$ and $M_3$ Weyl loop Fermi surfaces and we again directly observe from our photoemission spectra that the two loops link each other twice (Fig. \ref{Fig3}\panb). Repeating the analogous procedure for $X_3$, we observe that the $M_3$ and $M_1$ Weyl loops also link twice (Fig. \ref{Fig3}\panc). To estimate the stability of these links, we can measure how far apart one would need to slide two Weyl loops in order to unlink them (\exdas \ref{EFig2}, \ref{EFig3}). From the loop Fermi surfaces, we find that the typical `depth' of the link in momentum-space is $d_{\rm avg} = 0.58 \pm 0.08$ \AA$^{-1}$, of the same order of magnitude as the radius $|k|_{\rm avg}$ of the Weyl loop. This large depth suggests that the system lives well within a linked electronic phase. Our three-dimensional momentum-space analysis of the photoemission spectra suggests that each of the $M_1$, $M_2$ and $M_3$ Weyl loops links each other Weyl loop twice, forming a robust linked structure.

To further explore this link, we examine all three Weyl loops simultaneously using the experimentally-extracted loop trajectory, Eq. \ref{traj}. In an extended zone scheme, we plot the $M_1$, $M_2$ and $M_3$ Weyl loops around six nearby $X$ points, so that two redundant copies of each Weyl loop are included (Fig. \ref{Fig4}\panb). We find that the $M_1$ Weyl loop links both the $M_2$ and $M_3$ Weyl loops; the $M_2$ Weyl loop links both the $M_3$ and $M_1$ loops; and similarly for the $M_2$ loop. This suggests that the three Weyl loops together form a single composite linked structure. By plotting additional redundant copies of the loops in higher Brillouin zones, we can form a Weyl loop network extending outward to infinity. To more deeply explore this linked structure, we examine energy-momentum photoemission slices tangential to all three loops near their extrema (Fig. \ref{Fig4}\pana). All slices exhibit a cone dispersion, consistent with the Weyl loop electronic structure. Moreover, we find quantitative agreement between the Weyl loop extrema expected from Eq. \ref{traj} and the locations of the photoemission cone dispersions, for all three loops. This agreement again suggests the observation of a composite structure of three interwoven Weyl loops. To better visualize the complete link structure, we construct a link diagram for our Weyl loops. In such a link diagram, we flatten the link from three to two dimensions while preserving the link information using an over/under notation (illustrated for the example of a Hopf link, Fig. \ref{Fig4}\panc). Because the Weyl loop link lives in the periodic momentum space of the crystal, we flatten the link into the surface Brillouin zone. Moreover, because our analysis shows that all three Weyl loops are symmetry-related, we choose the (111) surface Brillouin zone (\exda \ref{EFigCrys}), which treats $X_1$, $X_2$ and $X_3$ equivalently (Fig. \ref{Fig4}\pand). The resulting link diagram shows three loops straddling the edges of the surface Brillouin zone (Fig. \ref{Fig4}\pane). We observe that the link wraps around $\mathbb{T}^3$ in all three momentum-space directions. This behavior suggests that the link is geometrically essential, in the sense that it cannot be smoothly perturbed to live entirely within a local region of the Brillouin zone. The link diagram further shows that each loop is linked with each other loop exactly twice. This gives $2$ for the geometric linking number, defined as the minimal number of crossing changes between link components needed to separate the components. The geometric linking number of the composite Weyl loop structure can then be written as $(2,2,2)$, where the first entry in the list corresponds to the linking number between $M_1$ and $M_2$, the second entry between $M_2$ and $M_3$, and the third entry between $M_3$ and $M_1$. By analogy with topological insulators and Weyl point semimetals, this Weyl loop link is expected to be stable under arbitrary, small, symmetry-preserving perturbations of the electronic structure.

Having systematically characterized the link structure in the bulk of \s, we next consider its topological surface states. Unlinked loop nodes host conventional drumhead surface states, which fill a simply connected region of momentum space in the surface Brillouin zone. By contrast, linked loops exhibit an alternating pattern of topologically distinct regions where surface states are either present or suppressed, and which are pinned together at generic points in momentum space. This topological structure is captured by the Seifert surface, defined as a three-dimensional surface which has the link as its boundary \cite{Seifert_1935}. For a condensed matter system, we consider a Seifert surface defined in $(k_x,k_y,k_z)$ and bounded by the linked loop nodes, with energy axis collapsed. For the minimal case of a Hopf link, the Seifert surface exhibits a branched structure which `wraps' around the link (Fig. \ref{EFig7}\pana, left). A two-dimensional projection of the Seifert surface then produces alternating filled and empty regions, which meet at characteristic touching points (Fig. \ref{EFig7}\pana, right). In the case of the Weyl loop link which we observe in \s, the Seifert projection on the (111) hexagonal surface Brillouin zone then predicts several alternating regions with and without topological boundary states (blue and white regions, Fig. \ref{EFig7}\panc), exhibiting touching points along $\bar{\Gamma}-\bar{K}$. Since the Seifert surface encodes the linking number \cite{Seifert_1935}, the topological boundary states are associated with a Seifert bulk-boundary correspondence. In this correspondence, the linking number of the Weyl loops in the bulk is encoded in a Seifert surface, whose projection gives the topological boundary states. A measurement of the bulk link determines the Seifert boundary states, while a measurement of the Seifert boundary states allows a reconstruction of the bulk linking number. To explore these possible surface states, we examine the (111) surface of our \s\ samples in \ab\ calculation and surface-sensitive vacuum ultraviolet (VUV) ARPES. On an energy-momentum cut through the touching point we observe in calculation a pair of surface modes pinned together at the Weyl loops (Figs. \ref{EFig7}\pand). Moreover, our photoemission spectra are consistent with our \ab\ prediction, suggesting the observation of Seifert boundary states approaching the Weyl loop linking point (Fig. \ref{EFig7}\pane). On iso-energy contours of the electronic structure, we expect to observe arc-like slices of the Seifert states, stretching across the filled regions and connecting the linked Weyl loops. Examining the Fermi surface obtained in calculation, we observe a sharp arc of surface states connecting the linked Weyl loops, consistent with the Seifert projection (Fig. \ref{EFig7}\panf, left). At the same time, the suppressed region exhibits no topological surface states in calculation. Our Fermi surface obtained by VUV-ARPES matches the \ab\ prediction well (Fig. \ref{EFig7}\panf, right). We observe distinct arcs of states connecting the linked Weyl loops across the topological region, corresponding to the topological surface states observed on the energy-momentum cuts (Figs. \ref{EFig7}\pand, \pane) and suggestive of Seifert states at the Fermi level in \s. Our \ab\ calculations and photoemission spectra suggest the observation of Seifert boundary states.

Our photoemission spectra, \ab\ calculations and theoretical analysis suggest the observation of a loop node link in a quantum magnet. On the sample surface, we further observe Seifert boundary states protected by the bulk link, indicating a Seifert bulk-boundary correspondence. These results establish a new bridge between physics and knot theory, motivating further exploration of links and knots in electronic structures. Moreover, the linked loop state in \s, as well as in other materials, may give rise to exotic response quantized to the linking number, such as a link-quantized topological magneto-electric effect \cite{WeylLink_Yee,Biao_WilsonLoops,ShouChengZhang_HelixNodalLine,Armitage_QuantizedKerr}. Since high-symmetry magnetic and correlated materials are abundant in nature, these ideas open the way to understanding the exotic behavior of a wide class of quantum magnets and superconductors, as well as their photonic analogs.


\section{Acknowledgments}

I.B. thanks Nikita Lvov and Zolt\'an Szab\'o for discussions on linking numbers. The authors thank D. Lu and M. Hashimoto at Beamline 5-2 of the Stanford Synchrotron Radiation Lightsource (SSRL) at the SLAC National Accelerator Laboratory, CA, USA for support. I.B. and D.M. thank Takayuki Muro for experimental support during preliminary ARPES measurements carried out at BL25SU of SPring-8 in Hyogo, Japan. I.B. thanks Biao Lian for discussions on the topological magneto-electric effect. I.B., T.A.C., X.P.Y. and D.M. thank Jessica McChesney and Fanny Rodolakis for experimental support during preliminary ARPES measurements carried out at BL29 of the Advanced Photon Source (APS) in Illinois, USA. I.B. acknowledges discussions with Boris Belopolski on Savitzky-Golay analysis. G.C. acknowledges the support of the National Research Foundation, Singapore under its NRF Fellowship Award (NRF-NRFF13-2021-0010) and the Nanyang Assistant Professorship grant from Nanyang Technological University. T.A.C. acknowledges support by the National Science Foundation Graduate Research Fellowship Program under Grant No. DGE-1656466. A.C. acknowledges funding from the Swiss National Science Foundation under Grant No. 200021-165529. The authors acknowledge synchrotron radiation beamtime at the ADRESS beamline of the Swiss Light Source of the Paul Scherrer Institut in Villigen, Switzerland under Proposals 20170898, 20190740 and 20191674. The authors further acknowledge use of Princeton’s Imaging and Analysis Center (IAC), which is partially supported by the Princeton Center for Complex Materials (PCCM), a National Science Foundation (NSF) Materials Research Science and Engineering Center (MRSEC; DMR-2011750). This research used resources of the Advanced Photon Source, a U.S. Department of Energy (DOE) Office of Science User Facility operated for the DOE Office of Science by Argonne National Laboratory under Contract No. DE-AC02-06CH11357. The authors acknowledge beamtime at BL25SU of SPring-8 under Proposal 2017A1669 and at BL29 of the APS under Proposals 54992 and 60811. K.M. and C.F. acknowledge financial support from the European Research Council (ERC) Advanced Grant No. 742068 “TOP-MAT”. C.F. acknowledges the DFG through SFB 1143 (project ID. 247310070) and the Würzburg-Dresden Cluster of Excellence on Complexity and Topology in Quantum Matter ct.qmat (EXC2147, project ID. 39085490). M.Z.H. acknowledges visiting scientist support at Berkeley Lab (LBNL) during the early phases of this work. Work at Princeton University was supported by the Gordon and Betty Moore Foundation (Grants No. GBMF4547 and No. GBMF9461; M. Z. H.). The ARPES and theoretical work were supported by the United States Department of Energy (US DOE) under the Basic Energy Sciences programme (Grant No. DOE/BES DE-FG-02-05ER46200; M. Z. H.). Use of the Stanford Synchrotron Radiation Lightsource (SSRL), SLAC National Accelerator Laboratory, is supported by the U.S. Department of Energy, Office of Science, Office of Basic Energy Sciences, under Contract No. DE-AC02-76SF00515. The authors acknowledge MAX IV Laboratory for time on the BLOCH Beamline under Proposal 20210268. Research conducted at MAX IV, a Swedish national user facility, is supported by the Swedish Research council under contract 2018-07152, the Swedish Governmental Agency for Innovation Systems under contract 2018-04969, and Formas under contract 2019-02496.

\section{Author contributions}

I.B., G. Chang, T.A.C. and M.Z.H. initiated the project. I.B., T.A.C., Z.-J. C. and M.Z.H. acquired and analyzed ARPES spectra with help from X.P.Y., D.M., J.-X.Y., M.L., N.S. and S.S.Z. ARPES measurements were supported by N.B.M.S., A.C., C.P., B.T., M.L., J.A. and V.N.S. G. Chang performed the first-principles calculations. I.B. performed the $k \cdot p$ model analysis with help from G. Chang and S.-M.H. I.B. performed the linking number analysis with help from C.H. G. Cheng and N.Y. performed the scanning transmission electron microscopy measurements. K.M., C.S. and C.F. synthesized and characterized the single crystals. I.B. wrote the manuscript with contributions from all authors.

\section{Competing Interests}

The authors declare no competing interests.

\section{Methods}

{\it Single crystal growth:} \s\ single crystals were grown using the Bridgman-Stockbarger method. A polycrystalline ingot was first prepared using an induction melt technique, with a stoichiometric mixture of Co, Mn and Ga metal pieces of 99.99\% purity. Then the powdered material was poured into an alumina crucible and sealed in a tantalum tube. Growth temperatures were controlled using a thermocouple attached to the bottom of the crucible. During the heating cycle, the material was melted at temperatures above 1200$^{\circ}$C and then slowly cooled below 900$^{\circ}$C.\\

\begin{figure}
\centering
\includegraphics[width=17cm,trim={0in 0in 0in 0in},clip]{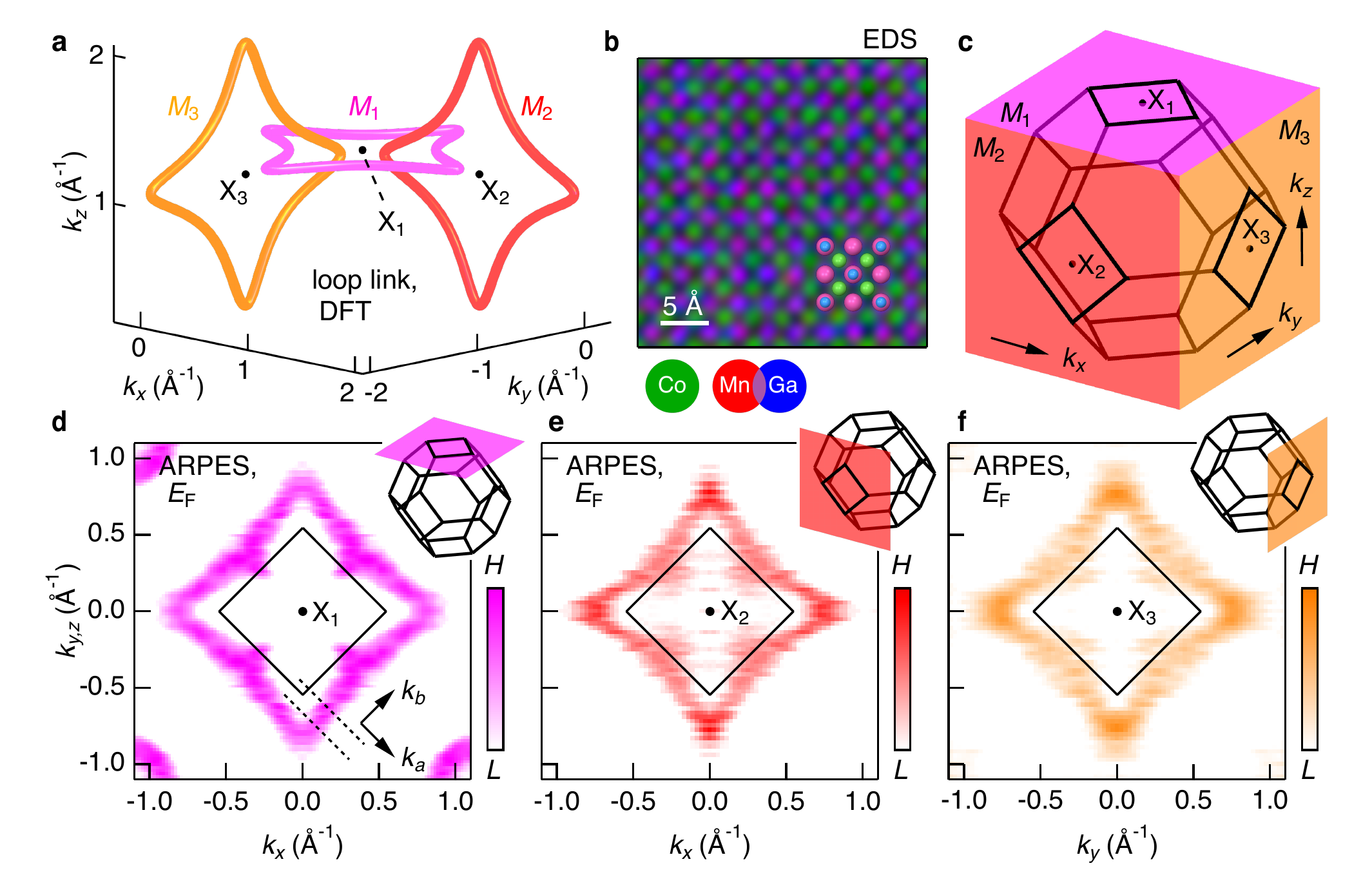}
\caption{\label{Fig1} {\bf Signatures of linked node loops in \s.} \cpana, Weyl loops in the electronic structure of \s, predicted by density functional theory (DFT). Three distinct Weyl loops are confined to the three mirror planes $M_1$, $M_2$ and $M_3$, in such a way that the loops link one another (additional copies of the loops in higher Brillouin zones not shown). \cpanb, Element-resolved crystal structure of \s\ along the [001] direction, acquired by atomic-level energy-dispersive X-ray spectroscopy (EDS). Atomic columns consist either entirely of cobalt (green) or alternating manganese (red) and gallium (blue). \cpanc, Bulk Brillouin zone (black truncated octahedron) of \s\ with three mirror planes indicated, $M_1$ (magenta, constant $k_z$), $M_2$ (red, constant $k_y$) and $M_3$ (gold, constant $k_x$). Each mirror plane contains square faces of the Brillouin zone. The high-symmetry momentum-space points at the center of each square are marked $X_1$, $X_2$, $X_3$. \cpand, Fermi surface acquired by angle-resolved photoemission spectroscopy (ARPES) at incident photon energy $544$ eV, corresponding to $M_1$. \cpane, Out-of-plane Fermi surface acquired on the same \s\ sample by an ARPES photon energy dependence from $500$ eV to $800$ eV in steps of $2$ eV, corresponding to $M_2$. \cpanf, Analogous out-of-plane Fermi surface corresponding to $M_3$, again on the same sample.}
\end{figure}

\begin{figure}
\centering
\includegraphics[width=16cm,trim={0in 0.4in 0in 0in},clip]{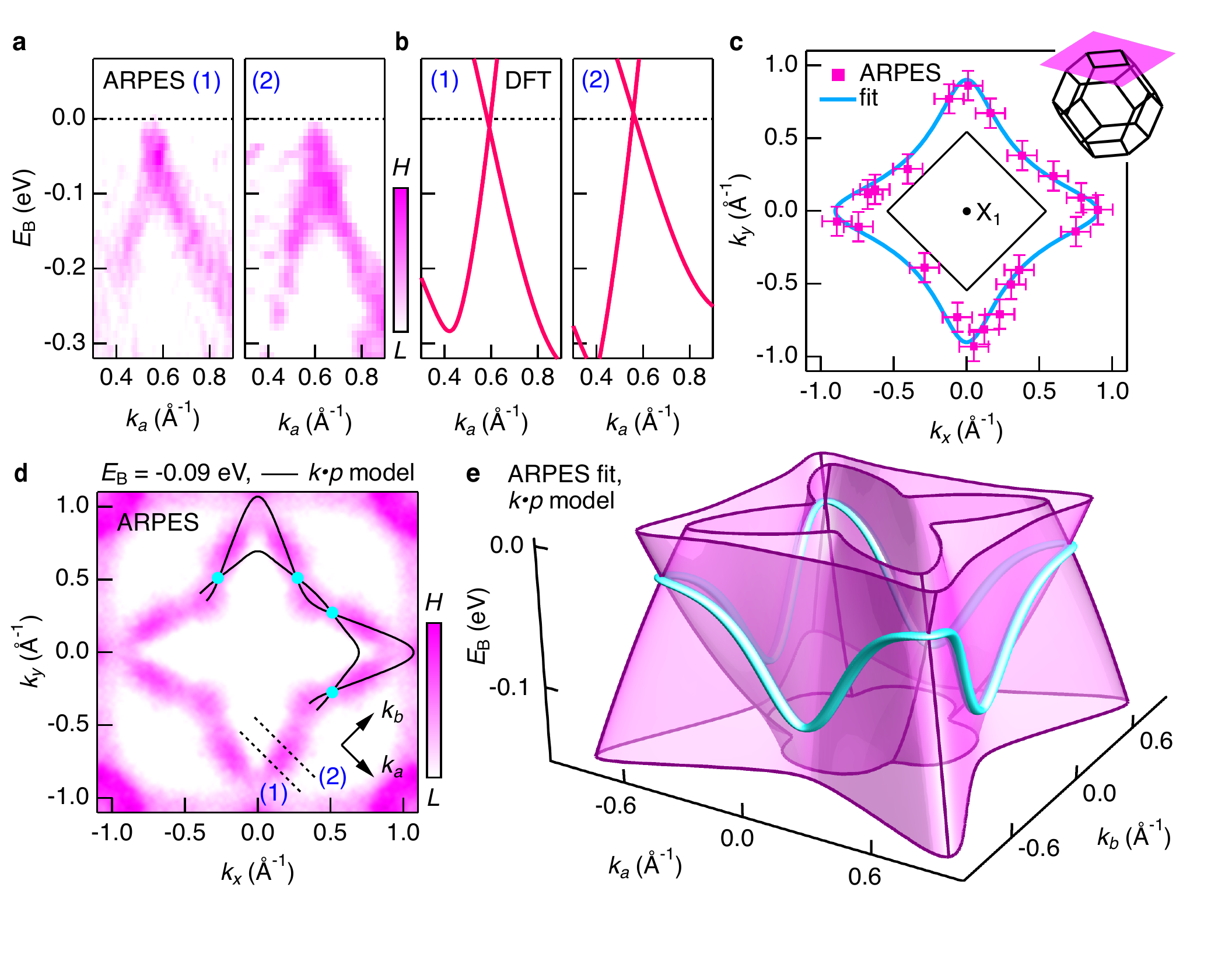}
\caption{\label{Fig2} {\bf Weyl loop trajectory in \s.} \cpana, Energy-momentum photoemission slices through the loop Fermi surface (slice locations marked by the dotted lines in (\pand) and Fig. \ref{Fig1}\pand). \cpanb, Energy-momentum slices through the Weyl loop from DFT, showing a Weyl loop cone (slice locations marked in \exda \ref{EFig1}\panc). \cpanc, Cone locations (magenta squares) systematically extracted from cone dispersions observed in photoemission spectra on $M_1$. Experimental loop trajectory extracted by fitting to the cone locations (cyan, see main text). The binding energy axis is collapsed. \cpand, Constant-energy photoemission slice with analytical model of the Weyl loop (black lines). This slice intersects the Weyl loop at a discrete set of points (cyan dots). \cpane, Dispersion of an effective $k \cdot p$ Hamiltonian for the Weyl loop, capturing the experimental loop trajectory.}
\end{figure}

\begin{figure}
\centering
\includegraphics[width=8cm,trim={0in 0in 0in 0in},clip]{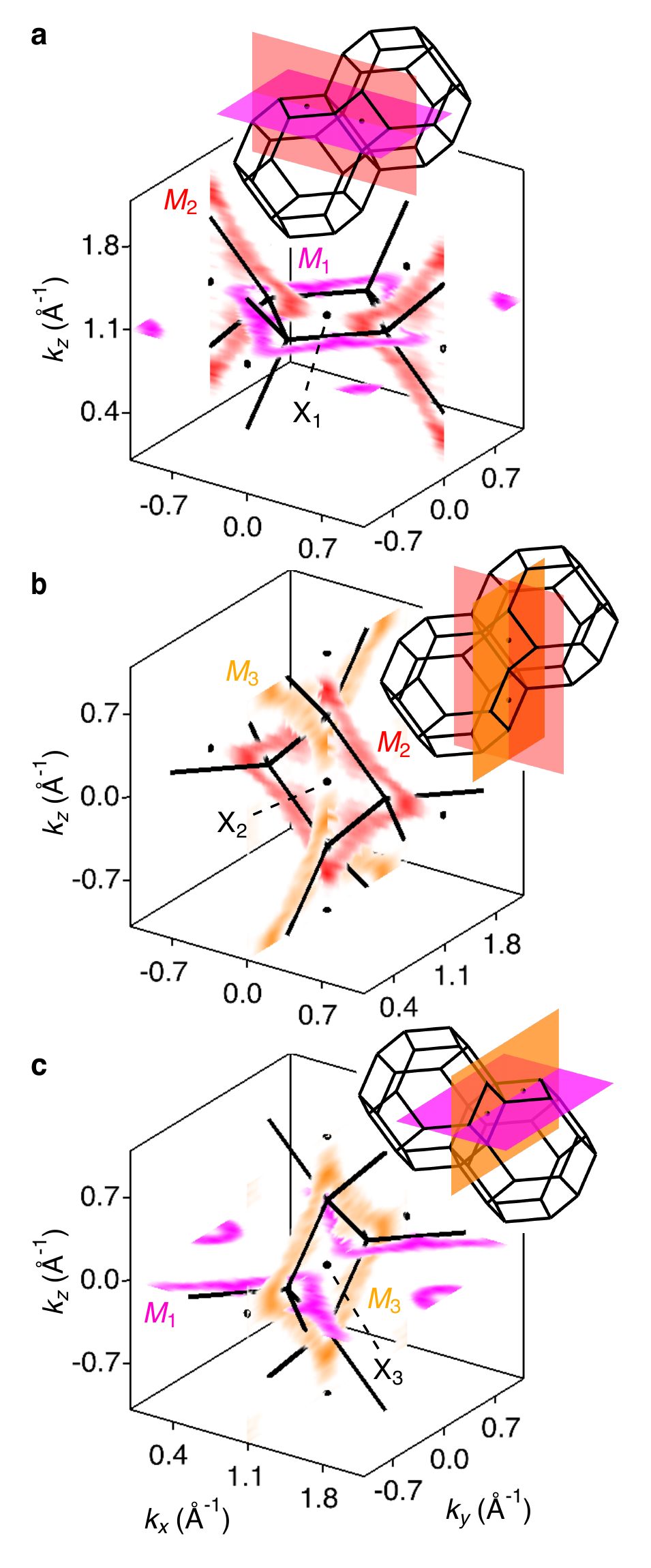}
\caption{\label{Fig3} {\bf Linked Weyl loops in \s.} \cpana, $M_1$ and $M_2$ loop Fermi surfaces from adjacent bulk Brillouin zones, plotted in an extended zone scheme, exhibiting a link structure. Inset: $M_1$ and $M_2$ plotted across multiple Brillouin zones. \cpanb, $M_2$ and $M_3$ loop Fermi surfaces from adjacent bulk Brillouin zones. \cpanc, $M_3$ and $M_1$ loop Fermi surfaces from adjacent bulk Brillouin zones.}
\end{figure}

\begin{figure}
\centering
\includegraphics[width=16cm,trim={0in 0.1in 0in 0in},clip]{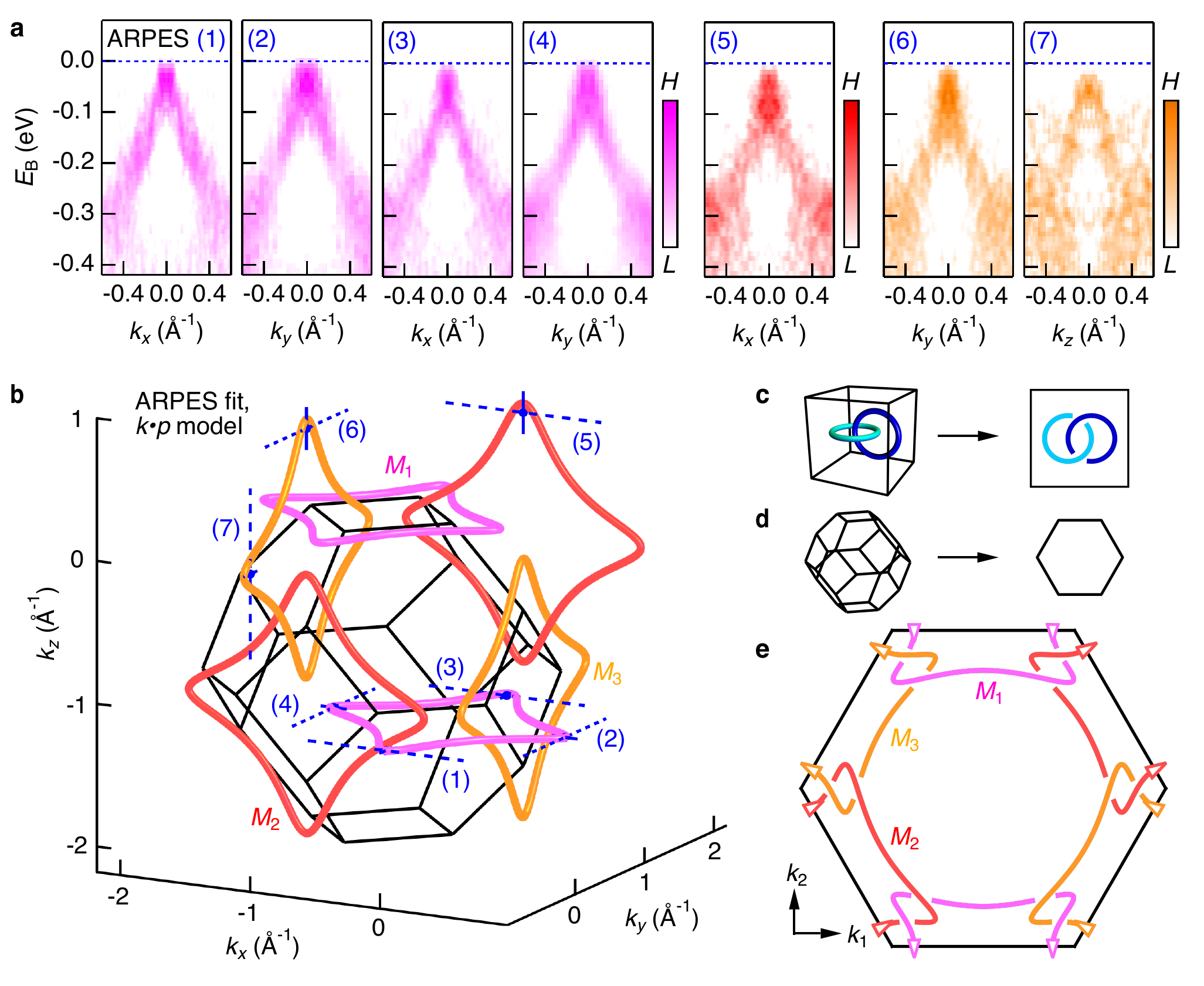}
\caption{\label{Fig4} {\bf Linking number $(2,2,2)$ in topological quantum matter.} \cpana, Energy-momentum photoemission slices tangential to the $M_1$, $M_2$ and $M_3$ Weyl loops at their extrema. \cpanb, Weyl loops from adjacent bulk Brillouin zones, based on the analytical model extracted in Eq. \ref{traj} and Fig. \ref{Fig2}, exhibiting links. Weyl cone positions (blue dots) extracted from the slices in Fig. \ref{Fig4}\pana\ (dotted blue lines), consistent with the analytical model (short blue line segments indicate the error). \cpanc, Link diagrams help visualize a three-dimensional link structure by flattening it to two dimensions while retaining the link information, illustrated for the example of a Hopf link. \cpand, In a crystal, it is natural to draw link diagrams in the surface Brillouin zone, such as the (001) surface Brillouin zone (hexagon, \exda \ref{EFigCrys}). \cpane, Link diagram for the \s\ Weyl loop link. There are three distinct Weyl loops and each Weyl loop links each other Weyl loop exactly twice, giving linking number $(2,2,2)$. The arrows indicate out-of-plane wrapping: as one follows the loop in the direction of the arrow, the loop wraps out of the page, exiting the Brillouin zone from the front and re-entering from the back, at the same time reconnecting at the opposite edge of the hexagon.}
\end{figure}

\begin{figure}
\centering
\includegraphics[width=13cm,trim={0in 0in 0in 0in},clip]{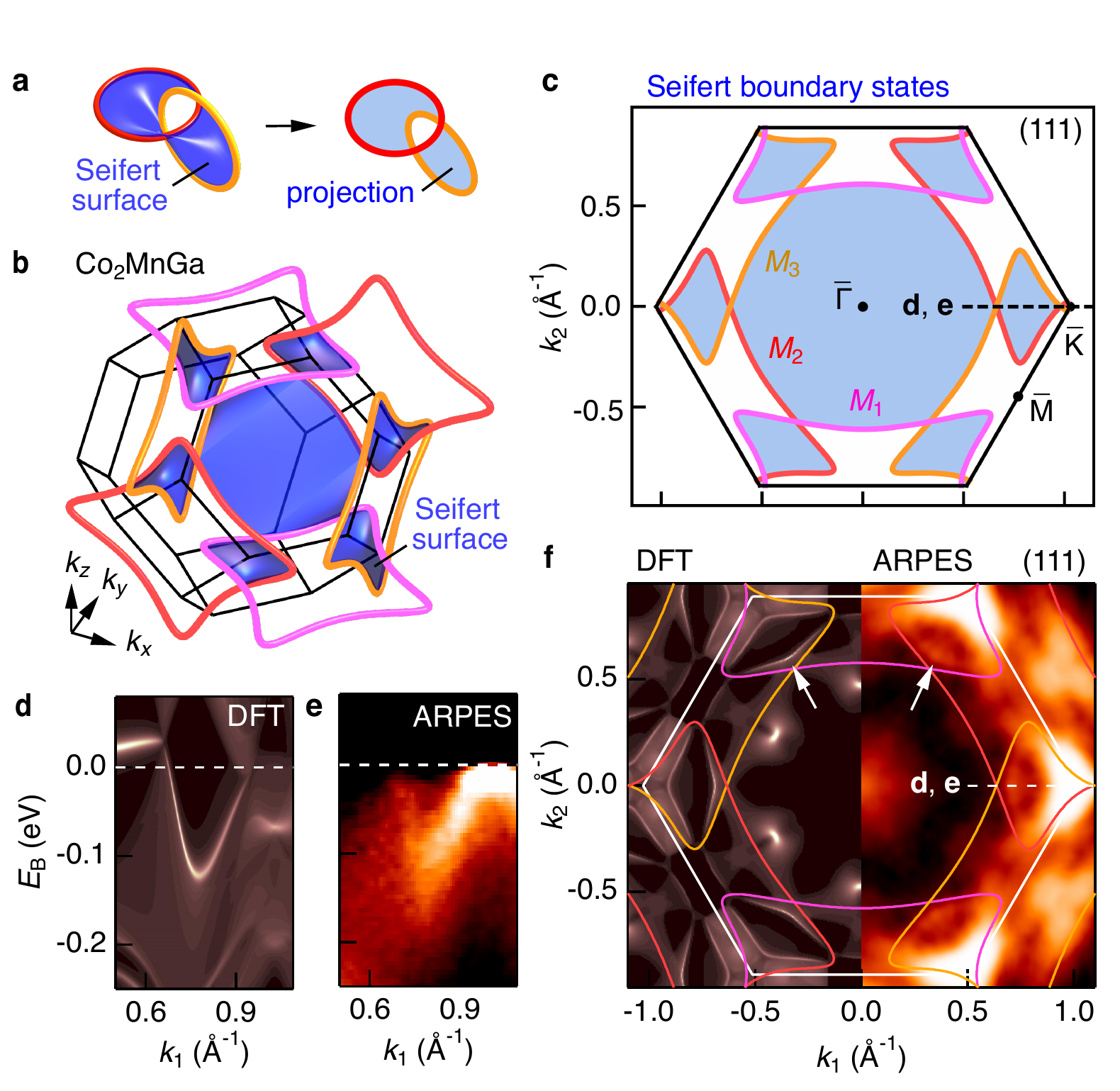}
\caption{\label{EFig7} \textbf{Seifert bulk-boundary correspondence.} \cpana, A Seifert surface is defined as a three-dimensional surface bounded by a link, shown for the example of a Hopf link. Its two-dimensional projection produces alternating filled and empty regions pinned together at characteristic touching points. \cpanb, In a condensed matter system, the Seifert surface is taken as a surface bounded by the linked loop nodes in three-dimensional momentum space $(k_x,k_y,k_z)$, shown for the case of the link observed in \s. \cpanc, The projection of the Seifert surface into the surface Brillouin zone is associated with topological boundary modes (blue regions) which touch at points in momentum space. Energy axis collapsed for clarity. \cpand, \Ab\ calculation of the surface states through the touching point, exhibiting pairs of boundary modes pinned together at the Weyl loops. \cpane, Surface-sensitive vacuum ultraviolet (VUV) ARPES energy-momentum cut through the touching point, exhibiting signatures of the pinned Seifert boundary modes, consistent with \ab\ calculations. Photon energy $h\nu = 63$ eV. \cpanf, Fermi surface in \ab\ calculation (left) and VUV-ARPES (right) exhibiting Seifert boundary modes that stretch across the topological regions, connecting different Weyl loops, consistent with the Seifert projection.}
\end{figure}

{\it Angle-resolved photoemission spectroscopy:} Soft X-ray ARPES measurements were carried out at the ADRESS beamline of the Swiss Light Source in Villigen, Switzerland under vacuum better than $5 \times 10^{-11}$ Torr and a temperature of 16 K \cite{ADRESS1_Strocov, ADRESS2_Strocov, ADRESS3_Strocov}. Rod-shaped single crystals of \s\ oriented along the conventional unit cell $\hat{z}$ direction were cleaved {\it in situ} at base temperature. The constant-energy cuts were symmetrized about $M_x$ and $M_{xy}$ (Fig.~\ref{Fig1}d), $M_x$ and $M_{xz}$ (Fig.~\ref{Fig1}e) and $M_y$ and $M_{yz}$ (Fig.~\ref{Fig1}f). The high-symmetry energy-momentum cuts were similarly symmetrized about $M_x$, $M_y$ or $M_z$, as appropriate and consistent with the nominal symmetries of the crystal (Fig.~\ref{Fig4}a). A background was removed from the photoemission spectra by a fixed intensity cutoff (raw, unsymmetrized data in \exdas~\ref{EFig6},~\ref{EFig5},~\ref{EFig9}). For the Fermi surfaces acquired at $h \nu = 544$ eV, the nominal energy resolution was $\delta E = 75$ meV; for the photon-energy dependences, the nominal energy resolution varied from $\delta E = 75$ meV at $h\nu = 500$ eV to $\delta E = 125$ meV at $h\nu = 800$ eV. The angular resolution was better than 0.2$^{\circ}$ in all cases. The Fermi surfaces were binned in an energy window of $\pm 38$ meV (Fig.~\ref{Fig1}d) and $\pm 25$ meV (Fig.~\ref{Fig1}e,f) around $E_{\rm F}$. Vacuum ultraviolet ARPES measurements were carried out at Beamline 5-2 of the Stanford Synchrotron Radiation Lightsource in Menlo Park, CA, USA at $\delta E = 15$ meV and temperature 20 K.\\

{\it \Ab\ calculations:} The electronic structure of \s\ in the ferromagnetic phase was calculated within the density functional theory (DFT) framework using the projector augmented wave method as implemented in the VASP package \cite{DFT2, DFT3}. The generalized gradient approximation (GGA) \cite{DFT4} and a $\Gamma$-centered $k$-point $12 \times 12 \times 12$ mesh were used. Ga $s, p$ orbitals and Mn, Co $d$ orbitals were used to generate a real space tight-binding model, from which Wannier functions were determined. The Fermi level in DFT was shifted to match the ARPES.\\

{\it Scanning transmission electron microscopy:} Thin lamellae for microstructure characterization were prepared from bulk single crystals by focused ion beam cutting using a ThermoFisher Helios NanoLab G3 UC DualBeam system (FIB/SEM). Atomic resolution high-angle annular dark-field (HAADF) scanning transmission electron microscopy (STEM) imaging and atomic-level energy-dispersive X-ray spectroscopy (EDS) mapping were performed on a double Cs-corrected ThermoFisher Titan Cubed Themis 300 scanning/transmission electron microscope (S/TEM) equipped with an X-FEG source operated at 300 kV with a Super-X EDS system.\\

\clearpage

\clearpage

\clearpage

\clearpage

\clearpage

\setcounter{figure}{0}
\renewcommand{\figurename}{Extended Data Fig.}

\clearpage
\begin{figure}
\centering
\includegraphics[width=10cm,trim={0in 0in 0in 0in},clip]{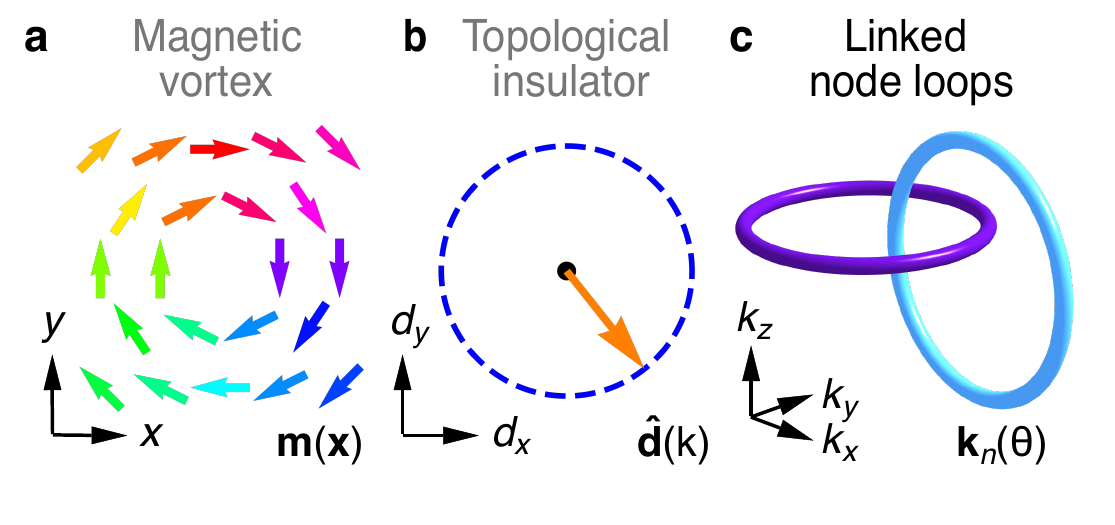}
\caption{\label{Fig0} {\bf Topological invariants in physics.} \cpana, An example of an order parameter winding in real space: a magnetic vortex. In this case, the order parameter is the local magnetization ${\bf m}({\bf x})$, confined to a magnetic easy plane in real space ($x, y$). It may happen that ${\bf m}({\bf x})$ winds around a point in real space, forming a magnetic vortex characterized by a winding number topological invariant, in this example given by $w = 1$. \cpanb, An example of a quantum wavefunction winding in momentum space: the one-dimensional topological insulator (Su-Schrieffer-Heeger model). This phase is described by Bloch Hamiltonian $h(k) = {\bf d}(k)\cdot \sigma$, where $k$ is the one-dimensional crystal momentum, $\sigma$ refers to the Pauli matrices and {\bf d}(k) is a two-component object confined to the ($d_x, d_y$) plane. The normalized quantity ${\bf \hat{d}}(k) \equiv {\bf d}(k)/|{\bf d}(k)|$ (orange arrow) moves around the unit circle (dotted blue) as $k$ varies. The topological invariant is related to how many times ${\bf \hat{d}}(k)$ winds around the origin as $k$ scans through the one-dimensional Brillouin zone. \cpanc, Node loops linking in momentum space: a three-dimensional electronic structure may exhibit multiple node loops (cyan and purple), characterized by ${\bf k}_n (\theta)$, where $n$ indexes the loops and $\theta$ parametrizes the loop trajectory in momentum space. The loops may link one another, encoding a linking number topological invariant. This example shows a Hopf link.}
\end{figure}

\clearpage
\begin{figure}
\centering
\includegraphics[width=14cm,trim={0in 0in 0in 0in},clip]{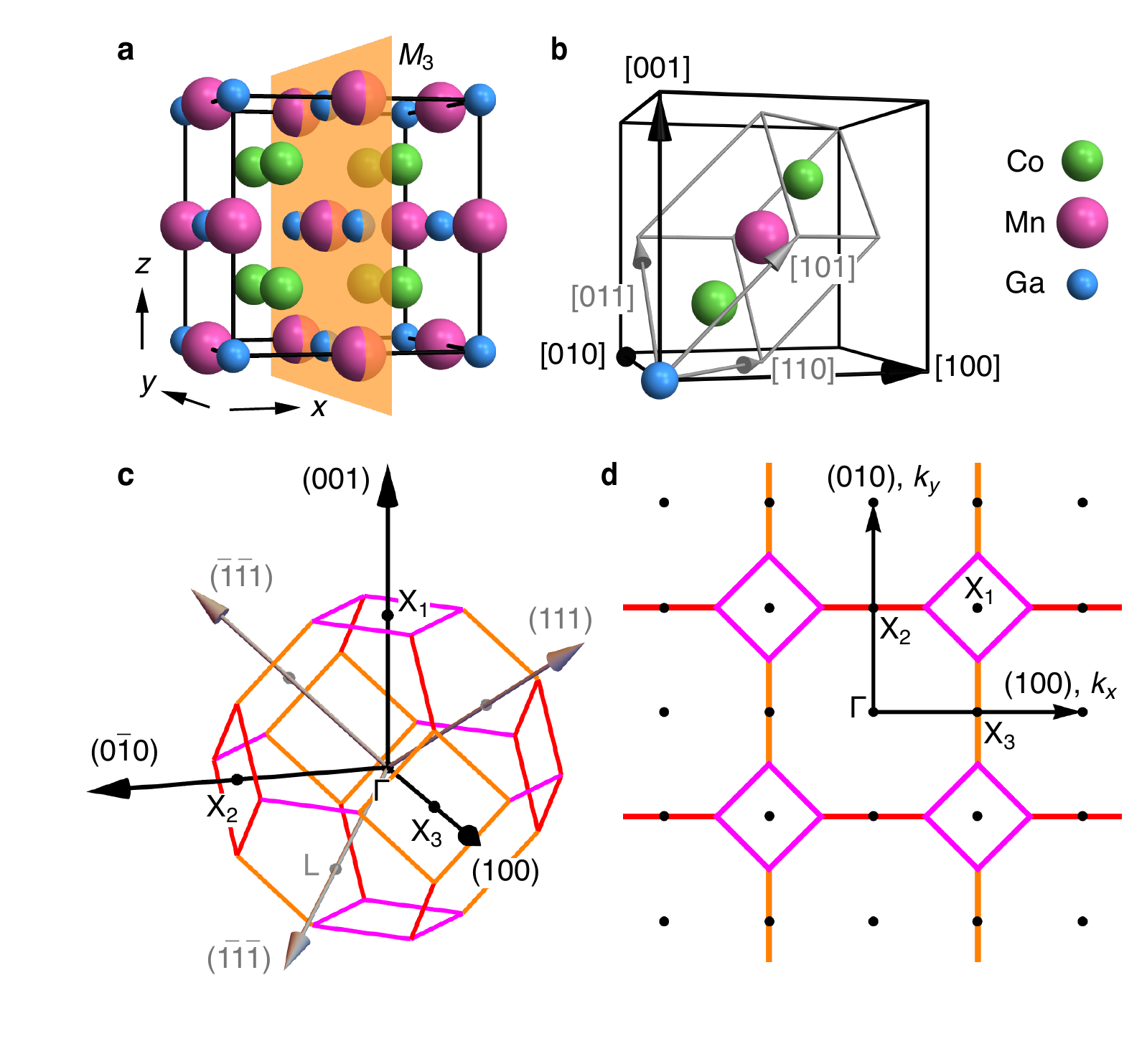}
\caption{\label{EFigCrys} \textbf{Crystal structure and Brillouin zone of \s.} \cpana, Conventional unit cell with representative crystallographic mirror plane $M$ (orange). \cpanb, The primitive unit cell (grey) includes one formula unit. \cpanc, Brillouin zone, with reciprocal lattice basis vectors (grey). In the reciprocal lattice basis, the $M_1$ plane corresponds to $(001)$, $M_2$ corresponds to $(010)$ and $M_3$ corresponds to $(100)$. \cpand, Slice through $\Gamma$ in an extended zone scheme.}
\end{figure}

\clearpage
\begin{figure}
\centering
\includegraphics[width=12cm,trim={0in 0in 0in 0in},clip]{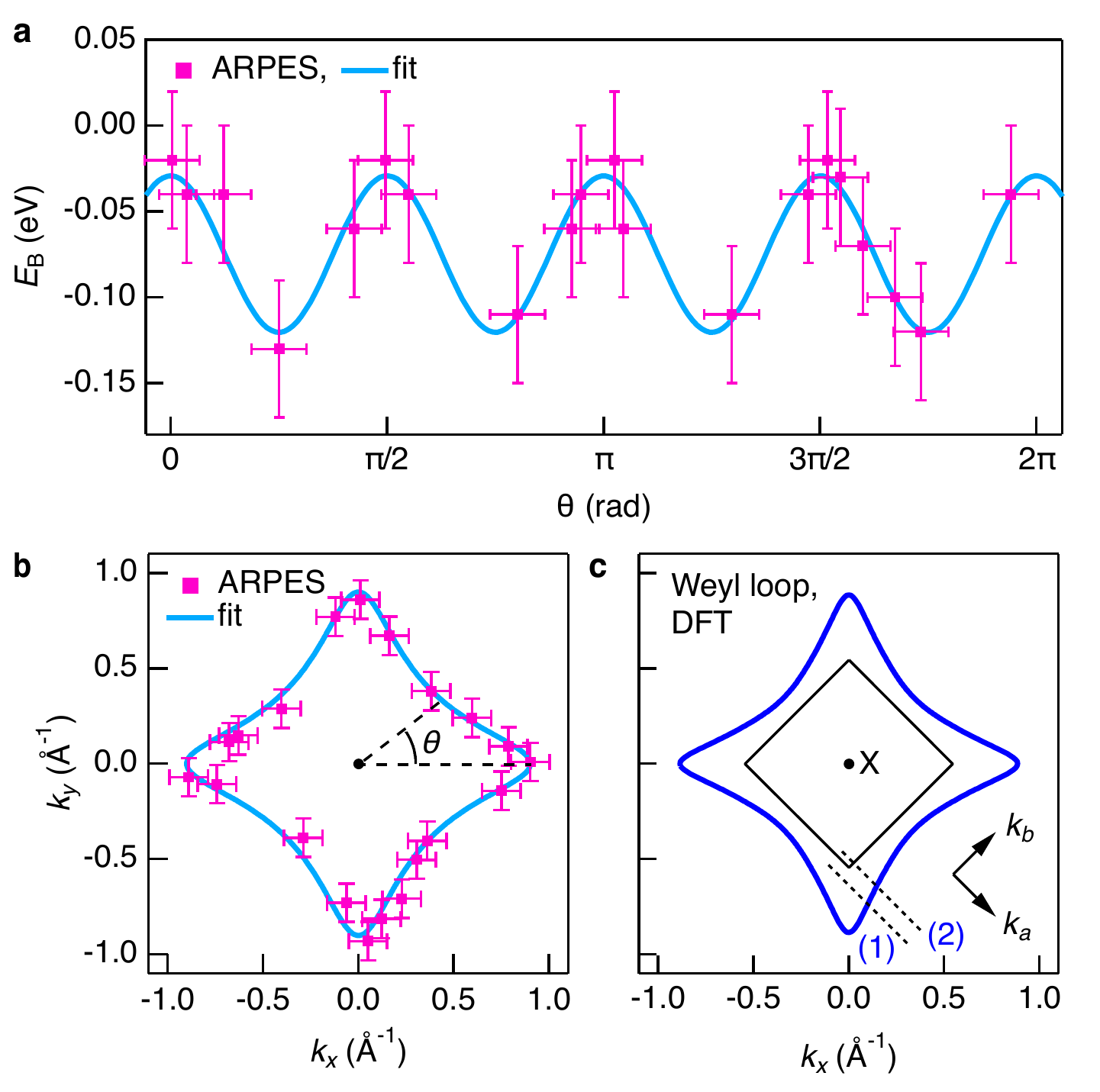}
\caption{\label{EFig1} \textbf{Energy dispersion of the Weyl loop.} \cpana, Crossing point energies $\eb$ and \cpanb, crossing point momenta $(k_x,k_y)$ systematically extracted from cone dispersions observed in the ARPES spectra (magenta squares), same dataset as Fig. \ref{Fig2}\panc\ ($h\nu = 544$ eV), with fit of the Weyl loop momentum trajectory and energy dispersion (cyan, see main text). The crossing point energies are parametrized by a polar angle $\theta$ defined by $\tan \theta \equiv k_y/k_x$. \cpanc, Weyl loop trajectory from DFT, with dotted lines indicating the DFT energy-momentum slices shown in Fig. \ref{Fig2}\panb. The binding energy axes in (\panb) and (\panc) are collapsed.}
\end{figure}

\clearpage
\begin{figure}
\centering
\includegraphics[width=17cm,trim={0in 0in 0in 0in},clip]{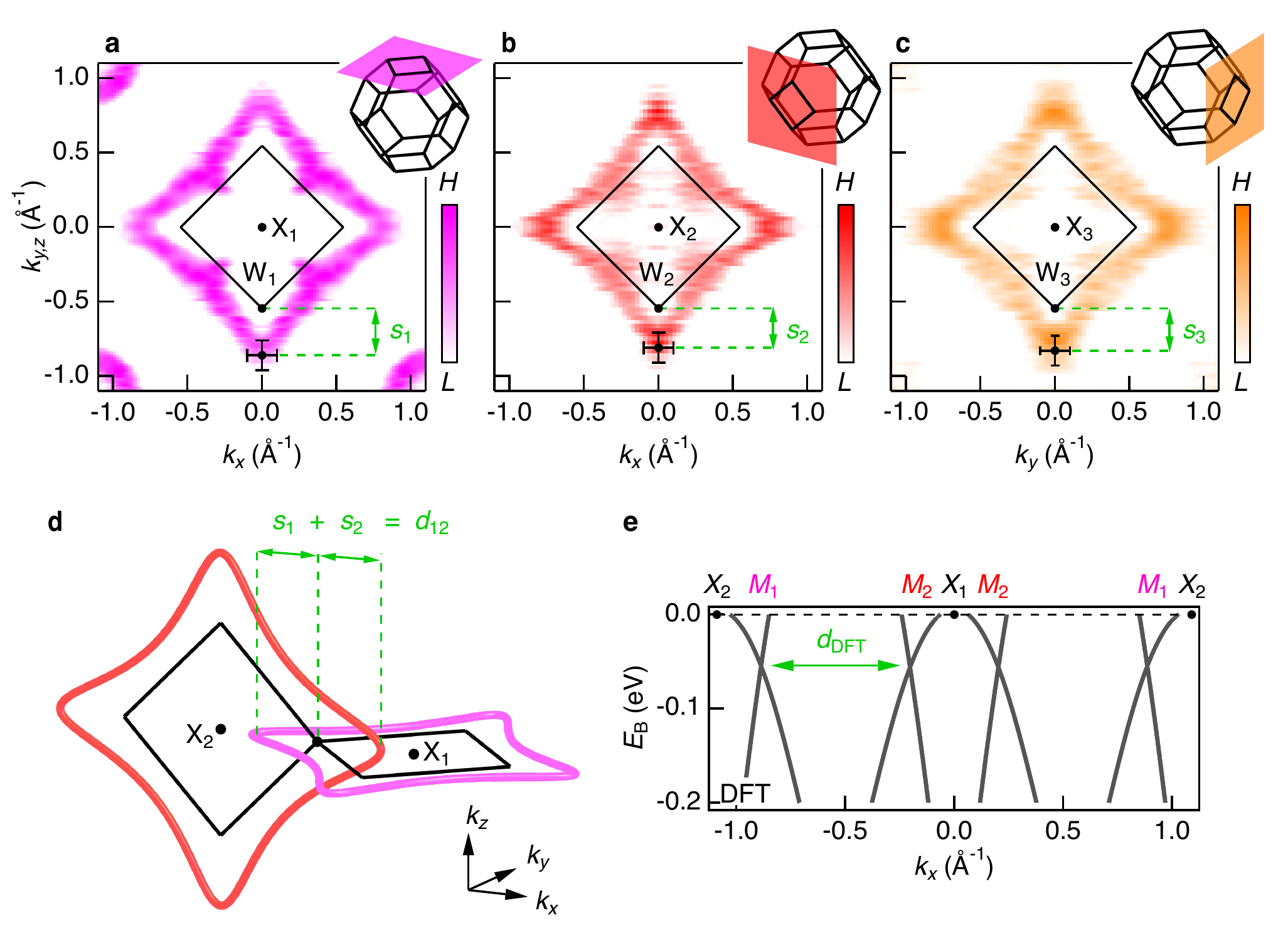}
\caption{\label{EFig2} \textbf{Link `depth' of the Weyl loops.} \cpana-\cpanc, Distance between the extrema of the Weyl loops and the bulk Brillouin zone $W$ points for the $M_1$, $M_2$ and $M_3$ Weyl loops. We estimate $s_1 = 0.32 \pm 0.1$ \AA$^{-1}$, $s_2 = 0.27 \pm 0.1$ \AA$^{-1}$ and $s_3 = 0.29 \pm 0.1$ \AA$^{-1}$. \cpand, The link depth captures how far in momentum space one would need to slide the Weyl loops in order to unlink them, providing a measure of the stability of the link. Based on the loop Fermi surfaces (\pana-\panc), we estimate $d_{12} = 0.58 \pm 0.14$ \AA$^{-1}$, $d_{23} = 0.55 \pm 0.14$ \AA$^{-1}$ and $d_{31} = 0.60 \pm 0.14$ \AA$^{-1}$. The average gives a typical link depth extracted from ARPES, $d_{\rm avg} = 0.58 \pm 0.08 $ \AA$^{-1}$. \cpane, Energy-momentum slice along the high-symmetry path $X_1-X_2$ from DFT, passing through two linked Weyl loops. We obtain $d_{\rm DFT} = 0.68$ \AA$^{-1}$.}
\end{figure}

\clearpage
\begin{figure}
\centering
\includegraphics[width=16cm,trim={0in 0in 0in 0in},clip]{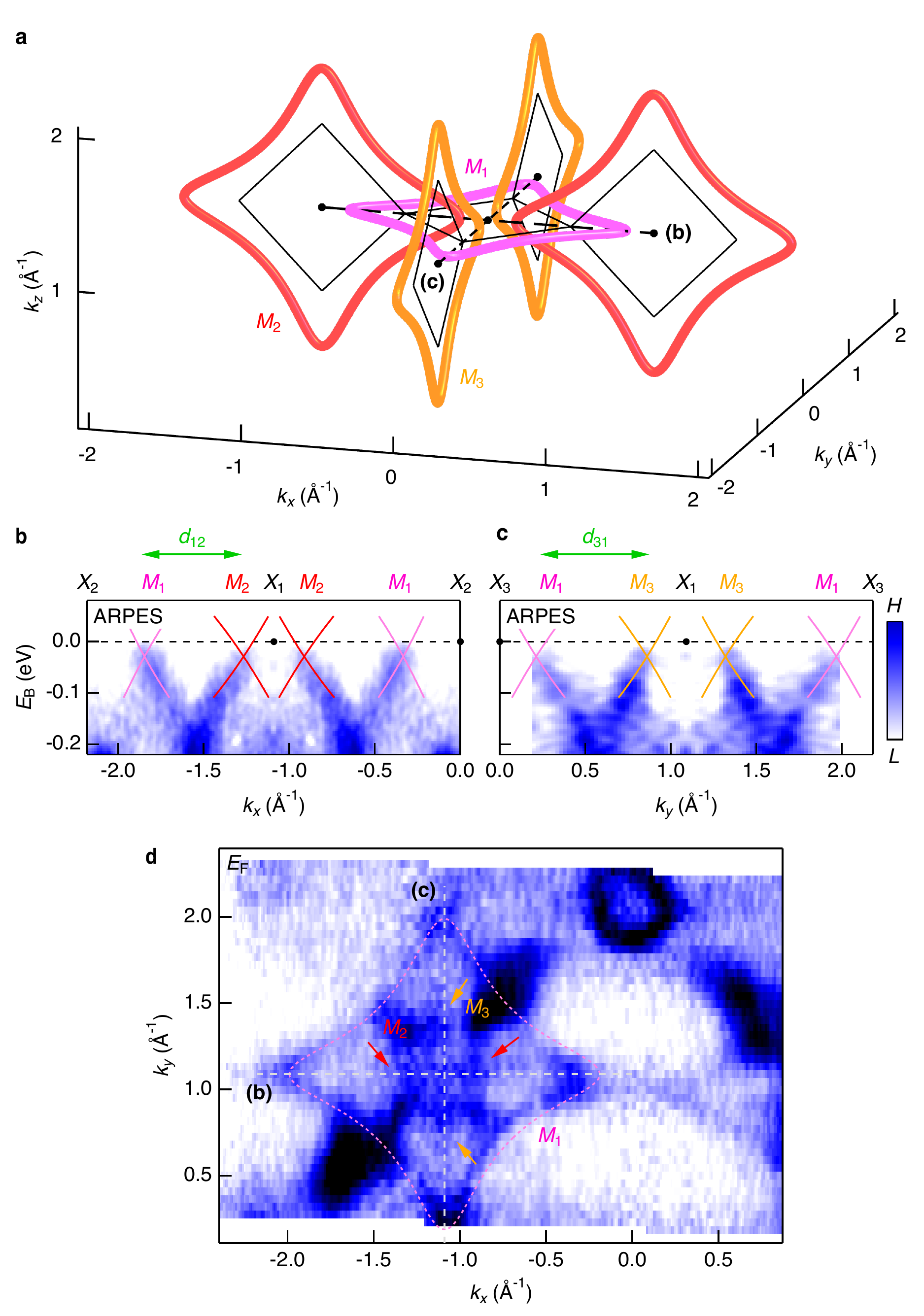}
\end{figure}

\newpage
\begin{figure}
\caption{\label{EFig3} \textbf{Supplementary measurement of the link depth.} \cpana, $M_1$, $M_2$ and $M_3$ Weyl loops, with trajectories obtained from the analytical model (see main text), showing that $M_1$ links $M_2$ twice and $M_3$ twice. Energy-momentum photoemission slices along the high-symmetry paths \cpanb, $X_1-X_2$ and \cpanc, $X_3-X_1$ obtained at photon energy $h\nu = 642$ eV. We observe $d_{12} = 0.56 \pm 0.1$ \AA$^{-1}$ and $d_{31} = 0.61 \pm 0.1$ \AA$^{-1}$, consistent with \exda \ref{EFig2}. \cpand, Fermi surface acquired at $h\nu = 642$ eV, exhibiting an in-plane Weyl loop contour, $M_1$. We further observe spectral weight emanating along $k_x$ and $k_y$ from the center of $M_1$, corresponding to the linearly dispersive branches in (\cpanb, \cpanc), again suggesting that $M_1$ is linked by $M_2$ and $M_3$.}
\end{figure}

\clearpage
\begin{figure}
\centering
\includegraphics[width=13cm,trim={0in 0in 0in 0in},clip]{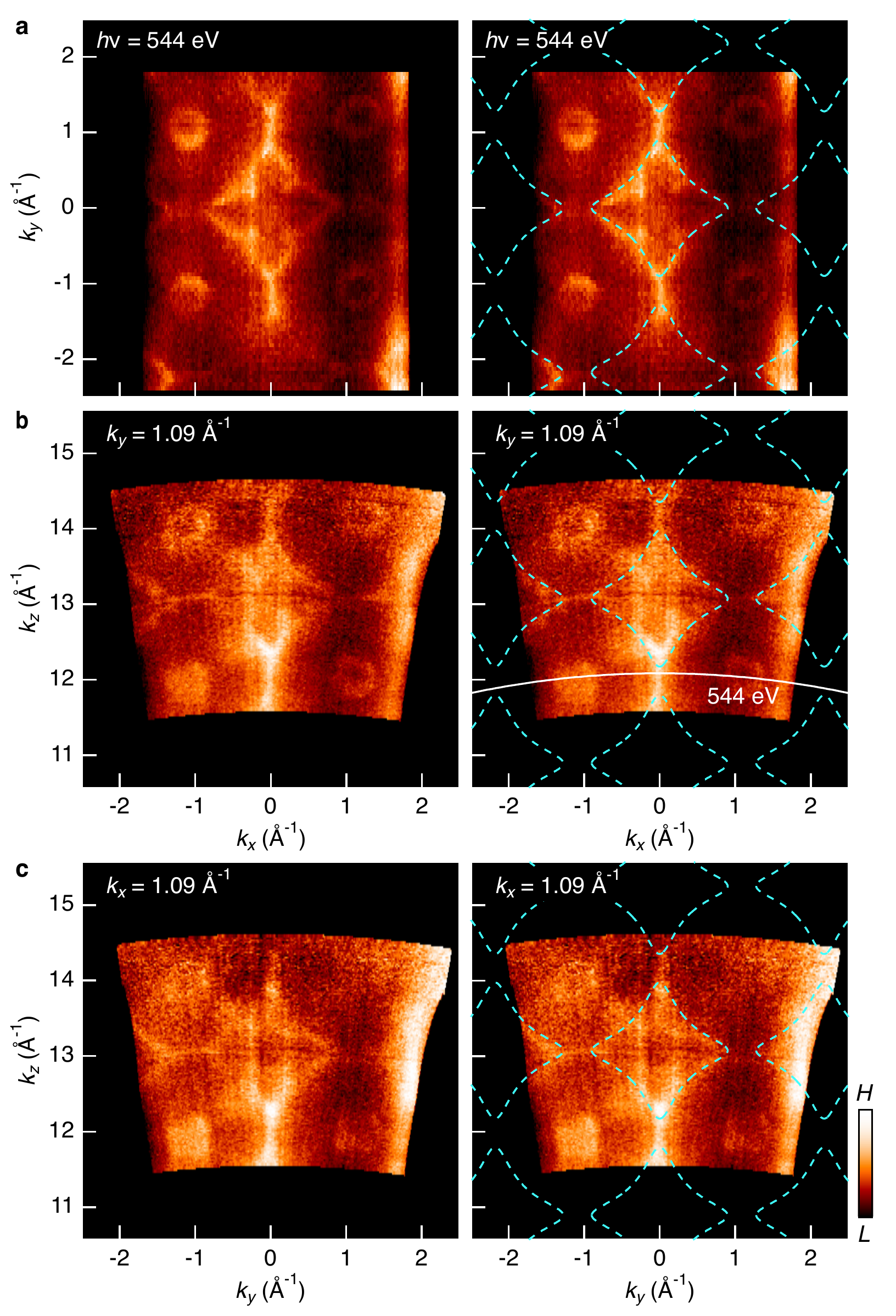}
\caption{\label{EFig6} \textbf{Unsymmetrized Fermi surfaces.} \cpana-\cpanc, Left: photoemission spectra displayed in Fig. \ref{Fig1}d-f, without symmetrization. Right: the same spectra, with the experimentally-determined Weyl loop trajectory overlaid across multiple Brillouin zones. The irrelevant $\Gamma$ pocket is consistently observed in all unsymmetrized spectra. Signatures of Weyl loops are observed around all $X$ points.}
\end{figure}

\clearpage
\begin{figure}
\centering
\includegraphics[width=16cm,trim={0in 0in 0in 0in},clip]{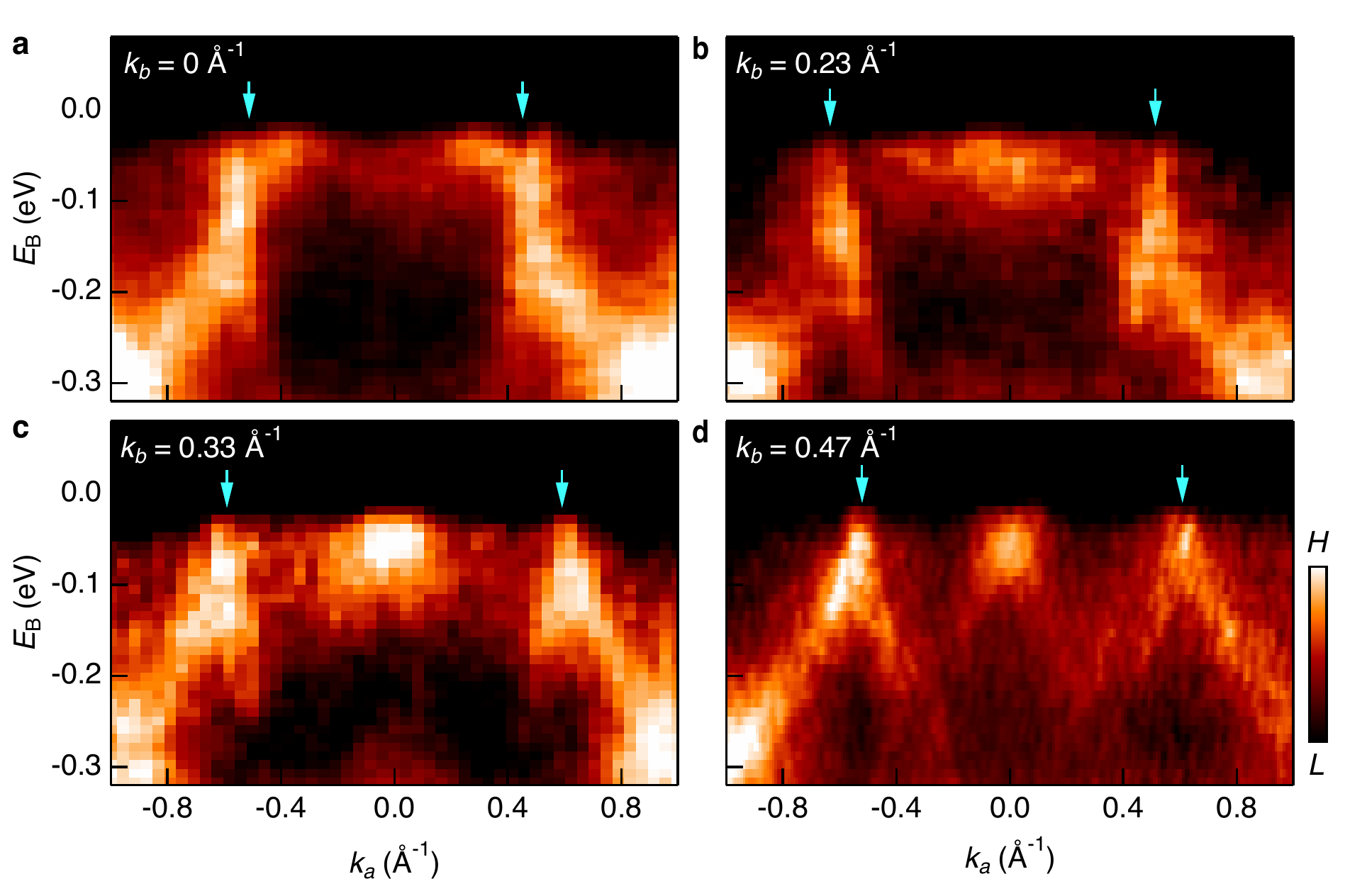}
\caption{\label{EFig5} \textbf{Energy-momentum cuts through the Weyl loop.} Photoemission spectra used to extract Fig. \ref{Fig2}c.}
\end{figure}

\clearpage
\begin{figure}
\centering
\includegraphics[width=16cm,trim={0in 0in 0in 0in},clip]{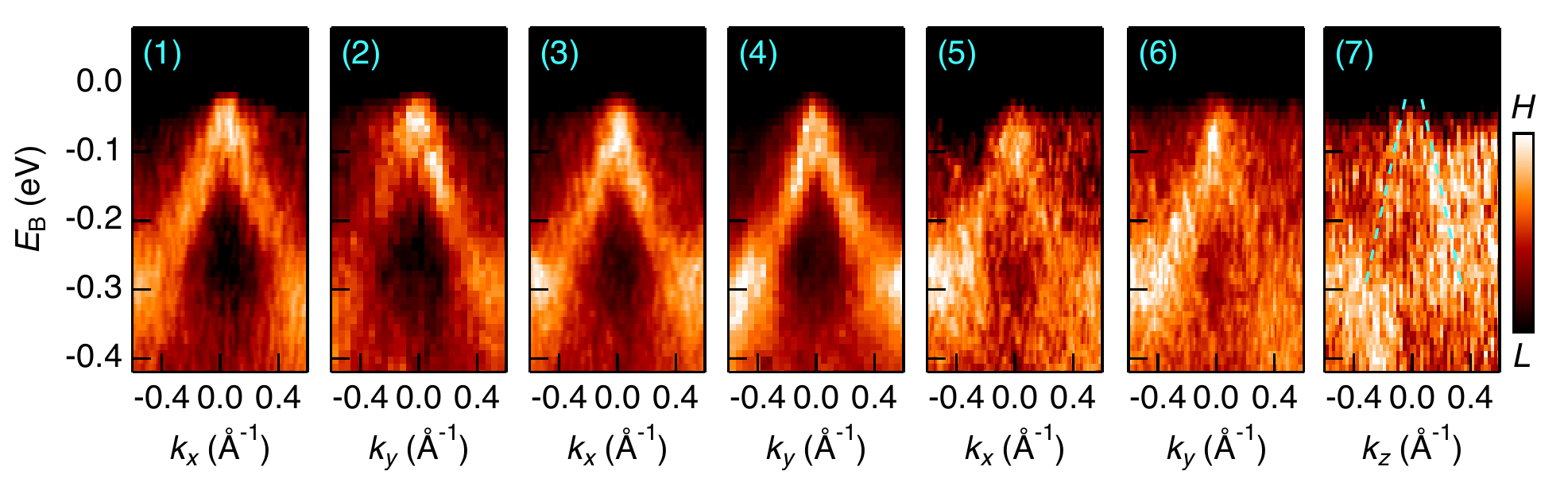}
\caption{\label{EFig9} \textbf{Unsymmetrized energy-momentum cuts.} Photoemission spectra displayed in Fig. \ref{Fig4}a, without symmetrization.}
\end{figure}

\clearpage
\begin{figure}
\centering
\includegraphics[width=14cm,trim={0in 0in 0in 0in},clip]{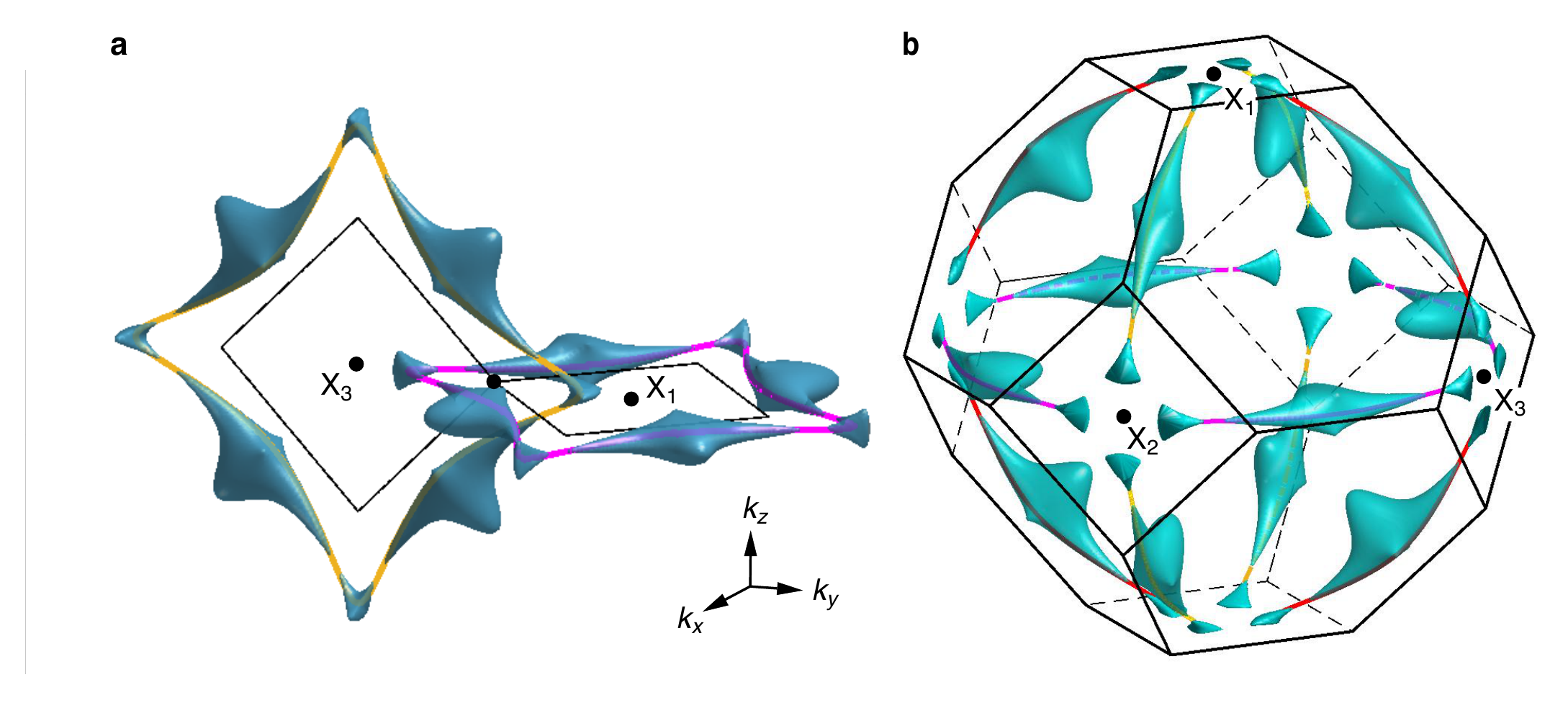}
\caption{\label{EFig4} \textbf{Linked Weyl loop Fermi surface.} Constant-energy slice of the pockets (navy) making up two linked Weyl loops obtained by \ab\ calculation, at binding energy $E_{\rm B} = -10$ meV, below the experimental Fermi level. The Fermi surface pockets touch at a set of discrete points, where the Weyl loop disperses through this particular $E_{\rm B}$. For reference, the full Weyl loop trajectories are indicated, collapsed in energy (orange around $X_3$, magenta around $X_1$). We observe that the Weyl loop Fermi surface pockets form a linked structure.}
\end{figure}

\clearpage
\begin{figure}
\centering
\includegraphics[width=16cm,trim={0in 0in 0in 0in},clip]{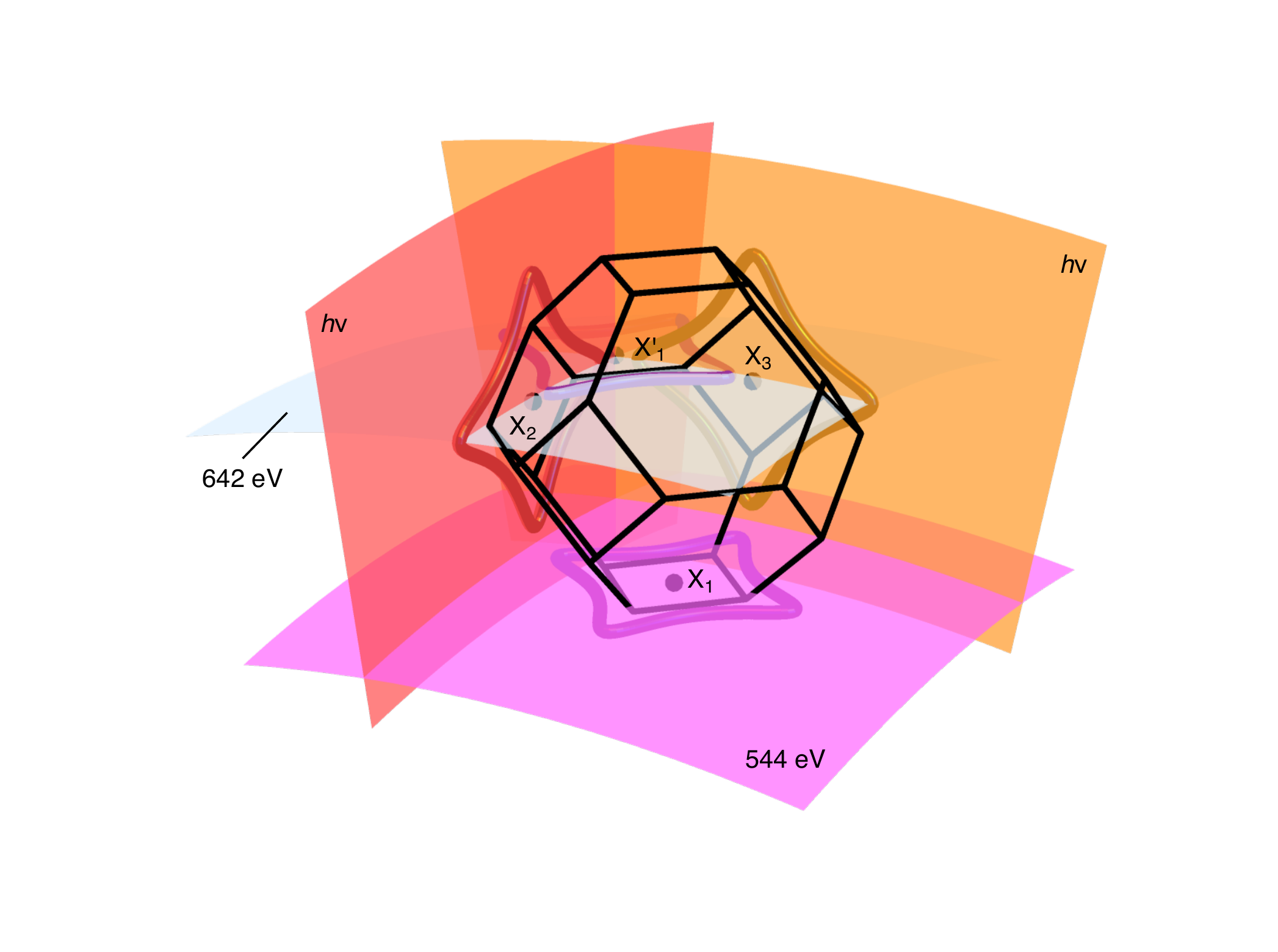}
\caption{\label{EFig8} \textbf{Measured Fermi surfaces in an extended zone scheme.} The Brillouin zone corresponds to $\Gamma_{(066)}$ in the primitive reciprocal basis.}
\end{figure}

\end{document}